\documentclass[showpacs,preprintnumbers,amsmath,amssymb]{elsart}

\usepackage{psfig}
\usepackage{epsf}
\usepackage{graphicx}
\usepackage{dcolumn}
\usepackage{bm}
\usepackage[figuresright]{rotfloat}

\begin{document}


\begin{frontmatter}

\title{CO adsorption on Cu(111) and Cu(001) surfaces: improving site preference in DFT calculations}

\author{Marek Gajdo\v s, J\"urgen Hafner}
\thanks[email]{E-mail: marek.gajdos@univie.ac.at}

\address{Institut f\"ur Materialphysik and
Center for Computational Materials Science\\
Universit\"at Wien, Sensengasse 8/12, A-1090 Wien, Austria
}

\begin{abstract}
CO adsorption on Cu(111) and Cu(001) surfaces has been studied within ab-initio density functional theory (DFT). The structural, vibrational and thermodynamic properties of the adsorbate-substrate complex have been calculated. Calculations within the generalized gradient approximation (GGA) predict adsorption in the threefold hollow on Cu(111) and in the bridge-site on Cu(001), instead of on-top as found experimentally. It is demonstrated that the correct site preference is achieved if the underestimation of the HOMO-LUMO gap of CO characteristic for DFT is correct by applying a molecular DFT+U approach. The DFT+U approach also produces good agreement with the experimentally measured adsorption energies, while introducing only small changes in the calculated geometrical and vibrational properties further improving agreement with experiment which is fair already at the GGA level.
\end{abstract}


\date{\today}

\end{frontmatter}

\vspace{0.2cm}
\section{Introduction}
\label{Introduction}

CO adsorption on transition metals has been studied already for half a century and is in general considered as being rather well understood. \cite{ss:Broden:59}. The models for chemisorption based on density functional theory (DFT) are generally accepted \cite{pr:Newns:178,rpp:Norskov:53} and it seems that there is nothing to investigate. However, there are some cases where DFT fails. The adsorption of a CO molecule on the Cu(111) surface is one of them  and the question why theory predicts adsorption at low coverage to take place in the hollow sites while experiments find the top site to be preferred is hence of general interest.

In a recent paper \cite{jpcm:Gajdos:16} we have tried to cover the state-of-the-art of the theoretical description of CO adsorption on close-packed surfaces of  transition and noble metals. Our calculated structural, vibrational and thermodynamical properties have been compared to available experimental data. The dependence of these basic properties on the filling of the $d$-band as well as the correlation with the experimental site-preference have been discussed. Although the choice of the exchange-correlation functional leads only to a small difference in the calculated structural and vibrational properties, it influences very strongly the absolute value of the adsorption energy \cite{jpcm:Gajdos:16}. A large effort has been devoted in the past to study  CO adsorption on Pt(111) \cite{jpcb:Feibelman:108} where CO also prefers to sit in the on-top site at low coverage, but where standard ab-initio calculations predict the hollow site again. Attempts to resolve this discrepancy have been undertaken by many researchers. Three ways to correct the prediction of the site preference have been proposed: (1) the molecular DFT+U method proposed by Kresse et al. \cite{prb:Kresse:68}, (2) the use of hybrid functionals incorporating a fraction of exact (i.e. Hartree-Fock (HF)) exchange in the DFT functional \cite{a:Mason:0310688}, and (3) relativistic corrections \cite{prb:Philipsen:56}. The former two approaches are based on the fact that the interaction between the empty CO 2$\pi^{\star}$ orbitals and the metallic $d$-band is overestimated due to the tendency of DFT to underestimate the gap between the highest occupied molecular orbital (HOMO - 5$\sigma$) and the lowest unoccupied molecular orbital (LUMO - 2$\pi^\star$) of the CO molecule. Therefore the position of the 2$\pi^{\star}$ orbital is shifted to higher values by an additional on-site Coulomb interaction U added to the DFT Hamiltonian or a mixing of DFT- and HF-exchange also leading to an increased level splitting. In contrast, the third approach concentrates on relativistic effects. Relativistic effects are small for the CO molecule and therefore changes in the $d$-band of the Pt substrate are considered as essential for restoring agreement with experiment. It is therefore of interest to re-investigate the problem of CO adsorption on the much lighter metal Cu where the same discrepancy between theory and experiment has been found.

The literature on CO adsorption on Cu(111) and Cu(001) is not as extensive as for transition-metal surfaces like Pt or Pd. The adsorption of  CO on Cu(111) is well understood from the experimental point of view. At low coverage and T$~=$~95~K the terminal (top) site is preferred, giving a ($\sqrt{3}\times\sqrt{3}$) LEED pattern. At higher coverages, bridge and hollow sites are also populated \cite{ss:Kessler:67,ss:Raval:203}. A contraction of the first layer appears at  0.44 ML, leading to a (1.5$\times$1.5)R18$^{\circ}$ structure \cite{prb:Moler:54}. In contrast ab-initio theory proposes the hollow site as the favorite site, but the difference in the adsorption energies is quite small ($\le$ 0.1 eV/molecule) \cite{ss:Lopez:477,jpcm:Gajdos:16}. On the other hand, the heat of adsorption hardly  changes up to a coverage of 0.44 ML \cite{ss:Hollins:89}.

On the Cu(001) surface a c(2$\times2)$ structure of the CO overlayer was observed  of in low energy electron diffraction (LEED) studies \cite{prl:Andersson:43,ss:McConville:166,jcp:Tracy:56,prb:Heskett:32}. The CO molecule is adsorbed in on-top, but the adsorption induces a negative shift of the workfunction at saturation coverage \cite{pn:Gartland:6}. For CO adsorption on Cu(001) surfaces the theoretical cluster study by van Daelen et al. \cite{ss:Daelen:417} reports the correct adsorption site, but the adsorption energies are too high (top site, local functional: E$_{ads} = -1.56$ eV) or too low (top site nonlocal functional: E$_{ads} = -0.39$~ eV) compared to experiment. Lewis and Rappe \cite{jcp:Lewis:110} report for the same site (top) a lower adsorption energy, but still too high (in absolute value) compared to experiment (E$_{ads} = -1.239$~eV). Although Ge and King \cite{jcp:Ge:111,jcp:Ge:114} found that bonding at the bridge site is stronger by 80 meV than in the top site, and a barrier between the top and the bridge site which was not observed in a helium atom scattering experiment (HAS) \cite{ss:Graham:427,jcp:Graham:114}. Finally, we know of only one attempt to study CO on the Cu(001) surface including the generalized gradient approximation (GGA) for the exchange-correlation functional \cite{jcp:Favot:114}. The authors demonstrated that the  revised Perdew-Burke-Erzernhof (RPBE) functional \cite{prb:Perdew:46,prb:Hammer:59} (which has been adjusted to fit CO adsorption data) leads to better agreement with experiment, but the adsorption energy is still stronger by about 0.2 eV (or 35 \%) than the experimental value.

In this paper we present an investigation of the adsorption of CO on the copper (111) and (001) surfaces using the molecular DFT+U method proposed by Kresse et al. \cite{prb:Kresse:68}. As copper is much lighter than platinum, relativistic corrections are expected to be less important, while the correction of the HOMO--LUMO gap of CO should lead to the correct site-preference also in this problematic case. Section \ref{Methodology} reviews the basic DFT methodology and introduces the DFT+U method.  Sections \ref{Cu(111) surface} and \ref{Cu(001) surface} summarize the description of the structural, vibrational and thermodynamic properties of the adsorbate-substrate complex within the DFT. Section \ref{Improving site preference within DFT+U method} applies the molecular DFT+U method and demonstrates that it corrects the prediction of the site preference agreement and leads to adsorption energies in agreement with the experiments, while introducing only small changes in the geometric and vibrational properties. In the last section we discuss the advantages of the method and summarize all important points.

\section{Methodology}
\label{Methodology}

\subsection{DFT calculations}

The calculations presented in this study are performed within a plane-wave based  density functional framework. We have used the Vienna Ab-initio Simulation Package (VASP) \cite{vasp,prb:Kresse:54} and employed  projected augmented-wave (PAW) potentials \cite{prb:Blochl:50,prb:Kresse:59}. For the description of  exchange and correlation the LSD functional proposed by Perdew and Zunger \cite{prb:Perdew:23} is
used, adding (non-local) generalized gradient corrections (GGA) of various flavor (PW91 according to Perdew et al. \cite{prb:Perdew:46}, RPBE the Perdew-Burke-Ernzernhof functional as modified by Hammer et al. \cite{prb:Perdew:46,prb:Hammer:59}). The equilibrium lattice constants of Cu are $a_{\rm Cu} = 3.664$~\AA~(PW91), $a_{\rm Cu} = 3.669$~\AA~(RPBE), compared to an experimental value of $a_{\rm Cu} = 3.61$~\AA. In spite of small difference between the PW91 and RPBE lattice constant we have used the lattice constant calculated by PW91 for the calculations with RPBE. 

The substrates are modelled by four layers of metal separated by a vacuum layer of approximately the same thickness. The two uppermost substrate layers
and the CO molecule are allowed to relax. A c(2$\times$4) cell is used for the Cu(111) surface and a p(2$\times$2) cell for the  Cu(001) surface, resulting in a coverage of 0.25 monolayers.

The self-consistent densities are calculated by iterative diagonalization of the Kohn-Sham Hamiltonian. The Fermi population of the Kohn-Sham states is k$_B$T = 0.2 eV and all the energies have been extrapolated to k$_B {\rm T}\rightarrow$0 eV. We have varied the k-point mesh from (4$\times$3$\times$1) to (8$\times$6$\times$1) for the (111) surface and from (4$\times$4$\times$1) to (8$\times$8$\times$1) for the (001) surface. The results here are presented for the most dense k-point mesh. For all further computational details we refer to our previous paper on CO adsorption on transition and noble metals \cite{jpcm:Gajdos:16}.

\subsection{The molecular DFT+U method}

In the molecular DFT+U method \cite{prb:Kresse:68} the DFT functional is modified in such a way as to shift the LUMO to higher energies. The form of the total-energy functional is inspired by the conventional DFT+U method adding an on-site Coulomb repulsion to the DFT Hamiltonian in order to account for strong electronic correlation effects \cite{prb:Dudarev:57,prb:Liechtenstein:52},
\begin{equation}
E_{GGA+U} = E_{GGA} + \frac{U}{2}\sum_{\sigma = 1}^{2} \sum_{i = 1}^{2} (n_i^{\sigma} - n_i^{\sigma} n_i^{\sigma}).
\end{equation}
Here $E_{\rm GGA}$ and E$_{\rm GGA+U}$ are the standard DFT functional in the generalized gradient approximation and the modified density functional and $\sigma$ is a spin index. The $n_i^{\sigma}$ stands for the occupancies of the two LUMO (2$\pi^\star$) orbitals for up and down spin. The occupancies are determined by defining two projection operators which are 1 for the LUMO orbitals of the free CO molecule and 0 for all other molecular eigenstates. Within the projector-augmented wave (PAW) approach implemented in VASP the projection operators defining these occupancies are identified with the PAW projections operators acting in the augmentation spheres around the carbon and oxygen atoms only. The occupancies $n_i^{\sigma}$ are hence determined taking the sum of
\begin{equation}
n_i^{\sigma}[\Psi^{\sigma}] =  \large{\sum_{n,lm}}\langle \Psi^\sigma \mid \tilde{p}_{n,lm} \rangle \alpha^i_{n,lm}  \alpha^i_{n,lm} \langle \tilde{p}_{n,lm} \mid \Psi^\sigma \rangle 
\end{equation}
over all one-electron states $\Psi^\sigma$ with fixed spin $\sigma$. The $\tilde{p}_{n,lm}$ are the PAW projector functions for the states with quantum numbers $lm$ on atom n as defined in Ref.~\cite{prb:Kresse:59}. The real coefficients $\alpha^i_{n,lm}$ have to be chosen such that $n_i^{\sigma}$ is one for a one-electron state belonging to the $2\pi^\star$ LUMO of a gas-phase CO molecule and zero for all other states. If the molecular axis is oriented along z, the $2\pi^\star$ molecular orbital consists of two orbitals $2\pi_x^\star, ~2\pi_y^\star$. Hence the coefficients $\alpha_{{\rm O},p_x}$ and $\alpha_{{\rm C},p_x}$ have to be chosen such that the corresponding occupancies are zero for the $1\pi$ and one for the $2\pi^\star$ molecular orbitals, i.e.
\begin{equation}
n_{2\pi^\star_x}^{\sigma}[\Psi^\sigma_{2\pi^\star_x}] =1 ,~~~~~ n_{1\pi_x}^{\sigma}[\Psi^\sigma_{1\pi_x}] =1
\end{equation} where the $\Psi$´s stand for the $2\pi^\star_x$ and $1\pi_x$ one-electron states. These relations define two equations for the coefficients $\alpha_{{\rm O},p_x}$ and $\alpha_{{\rm C},p_x}$, and similar relations hold for the $\pi_y$ orbitals. For any further detail we refer to the paper by K\"ohler and Kresse \cite{prb:Koehler:prep}.

It is important to emphasize that this ansatz leaves all DFT total energies unchanged if the occupancies are zero or unity. Hence the bond-length, the vibrational eigenfrequencies and even the excitation energies of the free molecule remain entirely unchanged. The important point is that despite the unmodified ground-state energy and structure, the adsorption energy decreases substantially, and that, as shown in detail below, the site preference is corrected.

\begin{figure*}[htb]
\centerline{\psfig{figure=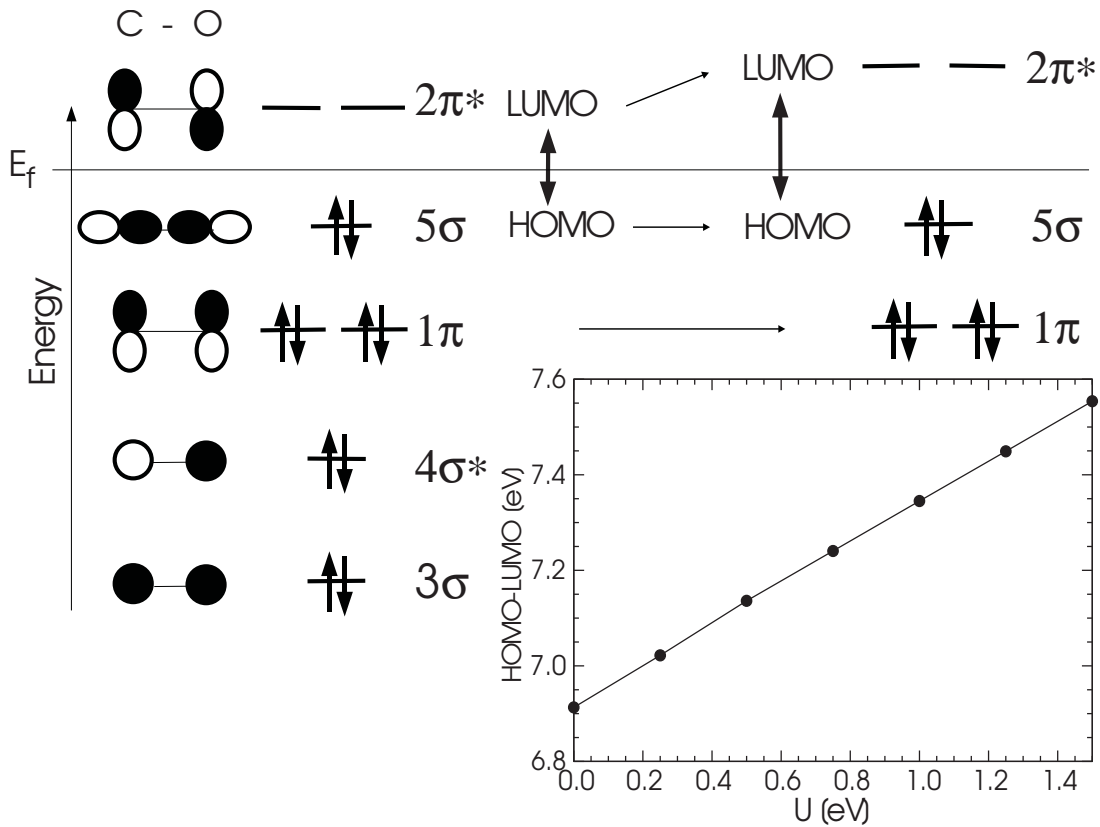,width=14.5cm,clip=true,angle=-90}}
\nopagebreak
\caption{Schematic sketch of the molecular eigenstates of the CO molecule. The DFT+U technique shifts the LUMO orbitals to higher energies, but the energies of the occupied orbitals remain the same. The inset shows the variation of the HOMO-LUMO for free CO with increasing U.}
\label{figure shift of LUMO}
\end{figure*}

Kresse et al. \cite{prb:Kresse:68} point out that with a modest value of $U =0.75$~eV, the DFT+U approach yields a single-particle $5\sigma \rightarrow 2\pi^\star$ gap in close agreement with the excitation energy for transferring a single electron from the $5\sigma$ to the $2\pi^\star$ state, while in the GGA without the Hubbard-like correction, the calculated excitation energy (not to be identified with the triplet to singlet excitation) is larger than the single-particle gap by 0.4 eV. This difference reflects the tendency of DFT to favor fractional orbital occupancies and hence to facilitate fractional electron transfer from the HOMO to the LUMO orbital. Other possibilities to correct this failure of DFT are hybrid functional incorporating a fraction of exact exchange or self-interaction corrected functionals.

\section{CO adsorption on Cu within DFT}

In this section we will briefly go trough our DFT results for CO adsorption on the Cu (111) and (001) surface. The structural, the vibrational and thermodynamic properties are presented.

\subsection{Cu(111) surface}
\label{Cu(111) surface}

In Table \ref{table CO on Cu(111)} the structural, vibrational properties and the  adsorption energies calculated using both the PW91 and RPBE functionals are presented.

\begin{sidewaystable}
\begin{tabular}{r|rrrr|r} 
 \hline 
  & \multicolumn{4}{c|}{theory} & experiment  \\
site  & top & bridge & fcc hollow  & hcp hollow  & top \\ 
\hline
E$_{ads}$ [eV/molecule]  &--0.73 [--0.42] (--0.43)& --0.74[--0.39](--0.25) &--0.84[--0.46](--0.27)&--0.83 [--0.45] (--0.26) & --0.43 to --0.52 \\
$d_{C-O}$ [\AA]         & 1.156 [1.162] (1.152)&  1.173 [1.179] &  1.180 [1.185] & 1.179 [1.185] & - \\
$h_{CO}$ [\AA]          & 1.96  [2.00] (1.98) &  1.55  [1.57]  &  1.43 [1.47]   & 1.43  [1.46]  & - \\
$d_{Cu-C}$ [\AA]        & 1.86  [1.87] (1.87)&  1.99  [2.01]  &  2.05 [2.08]   & 2.05 [2.08] & 1.91 $\pm$ 0.01 \\
$\Delta d_{12}$ [\%]    &--2.7  [--0.6] (--2.8) & --1.2  [1.6]   & --0.9  [1.0]   &--0.9  [1.1]& - \\
$b_{1}$ [\AA]           & 0.14  [0.17]  (0.14)&  0.11  [0.10]  &   0.11 [0.11]  &  0.09 [0.08] & - \\
$\Delta d_{23}$ [\%]    &--2.2  [--0.1] (--2.2)& --1.9  [--0.2] & --2.1  [0.2]   &--2.1  [0.1]& - \\
$b_{2}$ [\AA]           & 0.04  [0.04]  (0.04)&  0.05  [0.05]  &   0.03 [0.33]  &  0.07 [0.08] & - \\
\hline
$\nu_{S-CO}$ [cm$^{-1}$]&  322 [307] (308) &  281 [265]     &  290  [261]    &  289 [257] & 331-346 \\
$\nu_{C-O}$ [cm$^{-1}$] & 2046 [2034] (2075) & 1901 [1894]    & 1854 [1847]    & 1859 [1849] & 2071,2077\\
$\Delta\Phi$ [eV]             & 0.06 [0.09] (-0.22) &  0.58 [0.61]  &  0.72 [0.76] &  0.69 [0.74]& - \\
\hline 
\end{tabular}  
 \caption{Calculated structural, vibrational and thermodynamical properties of CO adsorbed 
 in the high symmetry sites on Cu(111) for a coverage  of $\Theta=\frac{1}{4}$~ML (c(2$\times$4)): $E_{ads}$ - adsorption energy, $d_{C-O}$ - carbon-oxygen bond length , $h_{CO}$ - height of CO molecule above the surface, $d_{Cu-C}$ - Cu-C bond length, $\Delta d_{12}$, $\Delta d_{23}$ - change of the average inter-layer spacing, $b_{1}$,$b_{2}$ - buckling of $1^{st}$ and $2^{nd}$ layer, $\nu_{C-O}$ - intramolecular stretching frequency, $\nu_{S-CO}$ - surface-adsorbate stretching frequency,  $\Delta \Phi$ - adsorption-induced change of the work-function. The results have been obtained with the PW91 functional, selected results calculated with the RPBE functional and PW91+U  method (U = 1.25 eV) are given in rectangular and round parentheses. Experimental data taken from Refs. \cite{cl:Vollmer:77,ss:Kirstein:176,ss:Kessler:67,jesrp:Hirschmugl:54,ss:Raval:203,cpl:Eve:313,jap:Michaelson:48} are listed where available.}
\label{table CO on Cu(111)}
\end{sidewaystable}

We discuss the PW91 results first. CO adsorption on the Cu(111) surface follows  the geometrical trends which depend on coordination \cite{jpcm:Gajdos:16}. The C-O bond is expanded compared to the gas-phase due to partial occupation of 2$\pi^{\star}$ orbital, the expansion increases with the number of Cu atoms to which the molecule binds. The Cu-C distance varies in the interval from 1.86 to 2.05 \AA~ and increases with coordination. The buckling of the first layer and the inward-relaxation of the top Cu-layer tend to decrease with coordination.

The experimental stretching frequency of the CO molecule ($\nu_{C-O}$) changes upon adsorption from the gas-phase value of $\nu_{C-O}$ = 2145 cm$^{-1}$ to about 2070 cm$^{-1}$ ( red-shift $\Delta\nu\sim$ --75 cm$^{-1}$), depending on coverage and temperature. The calculated stretching frequency of gas-phase CO frequency (2136 cm$^{-1}$, Ref.~ \cite{jpcm:Gajdos:16}) shifts to 2046 cm$^{-1}$ ($\Delta \nu \sim$ --90 cm$^{-1}$). In spite of the difference of 15 cm$^{-1}$ between the measured and calculated red-shifts, the agreement between experiment and theory is very satisfactory, because the stretching frequency of adsorbed CO decreases as the temperature decreases and therefore the shift measured in room-temperature experiments is smaller. The same trend (shift downwards) is observed for increasing coverage \cite{ss:Raval:203,ss:Hollins:89}.  The difference between the experimental and theoretical $\nu_{M-CO}$ is also quite small ( $\sim$ 10 cm$^{-1}$). Bridge and hollow adsorption lead to far larger redshifts of both the C--O and the substrate--CO stretching frequencies.

The workfunction of the clean surface is calculated to be $\Phi_{\rm clean}^{\rm Cu(111)}$= 4.78 eV for PW91 which is lower than experimentally estimated 4.98 eV \cite{jap:Michaelson:48}. On-top adsorption of CO leaves the work function almost unchanged ($\Delta \Phi = 0.06$~eV), while the stronger interaction in the sites with higher coordination leads to a pronounced increase of $\Phi$ by 0.6 eV (bridge) and 0.7 eV (hollow). 

Altogether we note good agreement of the vibrational frequencies  calculated for a linear adsorption geometry with experiment, while the too large adsorption energy and the too short Cu--C distance demonstrate that the overbinding tendency of the LDA is not completely removed by the PW91 form of the GGA functional. The most serious discrepancy, however, is the preference for binding in the hollow sites, the difference in the adsorption energy compared to the top site is 0.10 eV/molecule. Hammer et al. \cite{prb:Hammer:59} have noted that a slight modification of the PBE functional \cite{prb:Perdew:46} (which does not affect any of the sum rules) largely improves the prediction of the adsorption energies. This is confirmed by our calculations. Using the RPBE functionals we find adsorption energies in very good agreement with experiment (see Table~\ref{table CO on Cu(111)}). The energetic preference for adsorption in the fcc hollow is reduced to 0.03 eV/molecule, but still the wrong adsorption site is predicted. The change in the GGA functional has only a minor influence on the geometric and vibrational properties: the elongation of the C-O bond length is further increased by 0.006~\AA, the distance between the C-atom and the nearest Cu atom by 0.01~\AA~ (top) to 0.03~\AA~(hollow). The most important change occurs for the outermost interlayer distance of the Cu substrate which is expanded by about 3 \% compared to the PW91 results. The C-O stretching frequencies are reduced on average by about 10 cm$^{-1}$ for the gas-phase and adsorbed CO, so that the red-shift remains the same. A slightly more pronounced softening is observed for the adsorbate-substrate stretching frequency, but altogether the change of the GGA functional has only a small influence on geometry and dynamics.

\subsection{Cu(001) surface}
\label{Cu(001) surface}

For CO adsorption on this surface the number of theoretical investigations exceeds that of experimental studies - which could indicate that there are still unresolved questions. Our study covers adsorption in the three high-symmetry surface sites and the structural, vibrational and thermodynamic properties of CO on Cu(001) are compiled in Table \ref{table CO on Cu(001)}.

\begin{sidewaystable}
\begin{tabular}{r|rrr|r} 
 \hline 
 & \multicolumn{3}{c|}{theory} & experiment \\
  & top & bridge & hollow  & top\\ 
\hline
E$_{ads}$ [eV/molecule] & --0.863 [--0.565] (--0.567)&--0.883 [--0.545] (--0.430) &--0.842 [--0.471] (--0.189) &--0.53, --0.57 \\
\hline
$d_{C-O}$ [\AA]      & 1.156 [1.162] (1.152) & 1.170 [1.176] &  1.196 [1.200] & 1.13, 1.15 $\pm$ 0.1\\
$h_{CO}$ [\AA]       & 1.98  [1.99]  (2.00)  &  1.58 [1.61]  &  1.165 [1.20]  & - \\
$d_{Cu-C}$ [\AA]     & 1.85  [1.87]  (1.87)  &  1.99 [2.01]  &  2.15  [2.18]  & 1.90, 1.92$\pm$0.05 \\
\hline
$\Delta d_{12}$ [\%] &--4.1  [--1.2] (--4.3) &--2.5  [0.4]   &--1.6   [1.2]    & - \\
$b_{1}$ [\AA]        & 0.20  [0.19]  (0.19)  &  0.15 [0.14]  &  0.00  [0.00]  & - \\
$\Delta d_{23}$ [\%] &--1.5  [1.4]   (--1.5) &--1.2  [1.4]   &--6.7   [1.3]    & - \\
$b_{2}$ [\AA]        & 0.00  [0.00]  (0.00)  &  0.05 [0.05]  &  0.16  [0.15]  & - \\
\hline
$\nu_{S-CO}$ [cm$^{-1}$]&329 [313]  (310)  &  288  [272]   &  231   [206]    & 345 \\
$\nu_{C-O}$ [cm$^{-1}$]&2048 [2033] (2078) &  1926 [1917]  &  1727  [1729]  & 2079, 2085 \\
$\Delta\Phi$ [eV] & 0.29 [0.34] (0.03) & 0.65 [0.70]  & 0.89[0.94]  & 0.02, 0.26\\
\hline 
\end{tabular} 
 \caption{Calculated structural, vibrational and thermodynamical properties of CO adsorbed in 
 the high symmetry sites on Cu(001) for coverage $\Theta=\frac{1}{4}$~ML. The nomenclature is the same as in Table~\ref{table CO on Cu(111)}. Results obtained using the both the PW91, the RPBE (in rectangular parentheses) and the PW91+U (in round parentheses, U = 1.25 eV) functionals are listed. The experimental data are taken from Refs. \cite{cl:Vollmer:77,jcp:Tracy:56,prl:Andersson:43,ss:McConville:166,ss:Ryberg:114}.
}
\label{table CO on Cu(001)}
\end{sidewaystable}

If we compare CO adsorption on Cu(111) and Cu(001) similar trends are observed: the elongation of the C-O bond and the Cu-C bond-lengths depend on coordination. With increasing coordination the adsorbate-substrate bond length increases, the inward relaxation of the top layer and the buckling amplitude are reduced. Again we find a pronounced decrease of the adsorption energies if the PW91 functional is replaced by the RPBE functional. The reduction is stronger for the bridge and hollow sites, so that the RPBE calculation shows even a very slight preference for the experimentally observed on-top geometry by 0.02 eV compared to the bridge site. This energy difference is, however, much smaller that the 80 meV determined experimentally by Ge and King \cite{jcp:Ge:111,jcp:Ge:114}.

The adsorption geometries obtained with the different exchange correlation functionals differ only very slightly. The C-O bond length in the on-top configuration is 1.156 ${\rm\AA}$ for the PW91  functional, and 1.162 ${\rm\AA}$ for the RPBE functional. Similarly the metal-carbon distance is expanded from 1.85 ${\rm\AA}$ with PW91 to 1.87 ${\rm\AA}$ with RPBE. Agreement is slightly improved with the RPBE functional, but the difference is smaller than the experimental uncertainty.

The calculated red-shifts of the CO stretching mode are slightly smaller on the Cu(001) than on the Cu(111) surface for on-top ($\Delta \nu \sim -88~{\rm cm}^{-1}$) and bridge ($\Delta \nu \sim -210~{\rm cm}^{-1}$) adsorption. Again the calculation overestimates the redshift (experiment: $\Delta \nu \sim -66~{\rm cm}^{-1}$ for on-top CO). For adsorption in a hollow, however, one finds an increased red-shift of $\Delta \nu \sim$ --418 cm$^{-1}$, compared to $\Delta \nu \sim$ --286 cm$^{-1}$ on Cu(111),  reflecting the smaller adsorbate height in a four-fold than in a three-fold hollow and the larger stretch of the CO bond. The small changes of the frequencies calculated with the RPBE instead of the PW91 functional are of the same order as on the Cu(111) surface.

With $\Phi_{\rm clean}^{\rm Cu(001)}$ =  4.57 eV (PW91) and 4.49 eV (RPBE)  the workfunction of the clean Cu(001) surface is smaller than for the Cu(111) surface.  Experimentally values covering the interval from 4.59 to 4.83 eV have been reported \cite{pn:Gartland:6,b:Fauster:1995,jap:Haas:48}. Upon adsorption the workfunction increases quite strongly, depending on coordination of the adsorption site: using PW91 we find  $\Delta \Phi =$ 0.29 eV (top), 0.65 eV (bridge), and 0.89 eV (hollow), RPBE shifts are slightly larger by about 0.05 eV.

\section{Improving site preference within DFT+U method}
\label{Improving site preference within DFT+U method}

\begin{figure*}[htb]
\centerline{\psfig{figure=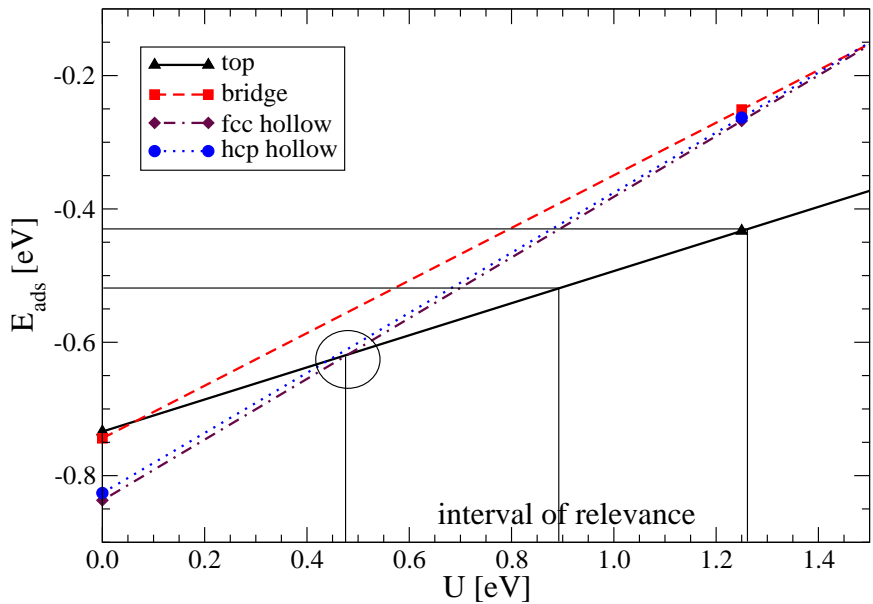,width=12cm,clip=true,angle=-90}}
\nopagebreak
\caption{ The adsorption energies of CO on the Cu(111) surface as a function of the  parameter U for the molecular DFT+U method. }
\label{figure CO adsorption with DFT+U 111}
\end{figure*}

\begin{figure*}[htb]
\centerline{\psfig{figure=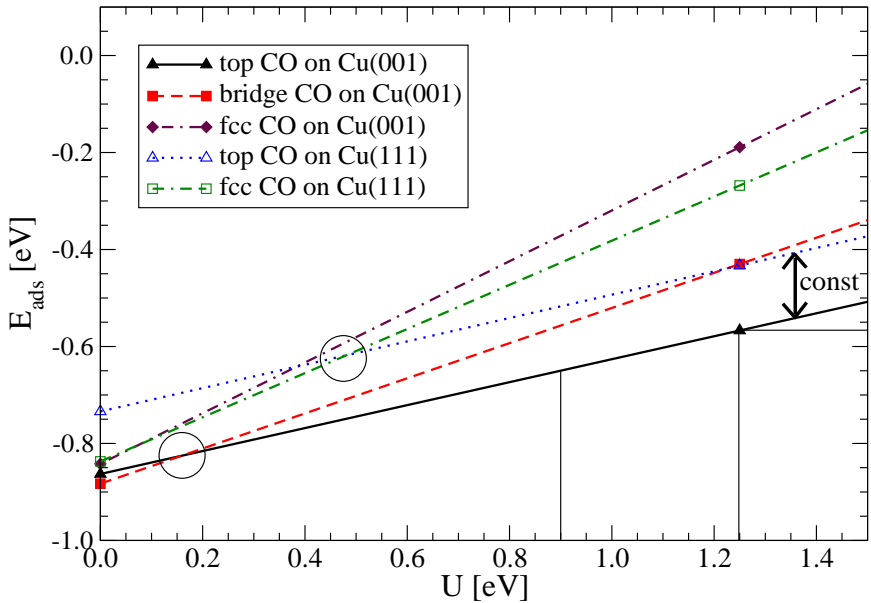,width=12cm,clip=true,angle=-90}}
\nopagebreak
\caption{ The adsorption energies of CO on the Cu(001) surface as a function  of the  parameter U in the molecular DFT+U method. }
\label{figure CO adsorption with DFT+U 001}
\end{figure*}

The DFT+U method enables us to study CO adsorption with different amounts of interaction between the unoccupied 2$\pi^{\star}$ molecular orbital and the metal d-band. We have studied the variation of the adsorption energy with an increasing value of the on-site potential U. Fig. \ref{figure CO adsorption with DFT+U 111} demonstrates the change in the adsorption energy of CO on the Cu(111) surface with increasing U. The calculations are based on the PW91 functionals - if the DFT+U approach is based on the RPBE functional, the adsorption energies are drastically underestimated. The adsorption energy linearly decreases with increasing U and the rate of reduction is greater for higher coordinated sites. Site-preference between top and bridge is reversed already around U=0.05 eV and between top and hollow around U=0.47 eV. For U between $\sim$ 0.9 eV and $\sim$ 1.25 eV the calculated adsorption energies fall into the interval covered by the experimental values. The same analysis for CO/Cu(001) is given in Fig.~\ref{figure CO adsorption with DFT+U 001}. Stability of on-top adsorption  over the bridge site is achieved at U$\ge 0.15$ eV, to match the experimentally determined difference in the adsorption energies ($\Delta {\rm E}_{ads} \sim$ 80 meV), a value of U$\sim$0.85 eV is required,  agreement with the experimentally measured adsorption energies is reached for U $\sim$ 1.25 eV. It is certainly encouraging the the same values of the on-site potential fit the CO adsorption data on both surfaces and that the optimal value of U determined from the adsorption study agrees reasonably well with that needed to bring the one-electron HOMO-LUMO gap in agreement with the excitation energies (cf. Sec.~\ref{Methodology}). A further confirmation that the molecular DFT+U approach grasps the correct physical mechanism comes from the observation that the difference of the on-top adsorption energies on the Cu(111) and Cu(001) surfaces is independent of U (see Fig.~\ref{figure CO adsorption with DFT+U 111}). For adsorption in the fourfold hollow on Cu(001), the decrease of the adsorption energy with U is faster than for adsorption in one of the threefold hollows on Cu(111), demonstrating the dependence of the $2\pi^\star$ backdonation on the coordination of the adsorbate. \\

The structural and vibrational properties of CO adsorbed in the top site on Cu(111) and Cu(001) calculated by the DFT+U approach are listed in Tables \ref{table CO on Cu(111)} and \ref{table CO on Cu(001)}. For these calculations we have used a Hubbard U parameter equal to 1.25 eV which is the highest acceptable value. The C-O bond (1.156 \AA~ calculated with PW91) is slightly shorter (1.152 \AA) and the Cu-C bond (1.85 \AA) is elongated to (1.87 \AA) for CO adsorption on Cu(111). The Cu--C bond length calculated by DFT+U has the same value as the calculation with RPBE functional (1.87 \AA). The structure of CO on Cu(001) is influenced in the same manner. As the DFT+U corrections affect only the adsorbate, it is not surprising that the calculated interlayer relaxations are the same as predicted using PW91 alone.

The stretching frequency of CO molecule ($\nu_{C-O}$) calculated using the DFT+U approach is shifted to higher frequencies by  20-40 cm$^{-1}$. The DFT+U frequencies of $\nu_{\rm C-O}$ are closer to the experimental values. The stretching frequency of CO ($\nu_{C-O}$) in top on Cu(111) is 2075 cm$^{-1}$ and the corresponding experimental values are 2071 and 2077 cm$^{-1}$. Similarly, the calculated $\mu_{C-O}$ in top on Cu(001) is 2078 cm$^{-1}$ and the experimentally observed values are 2079 and 2085 cm$^{-1}$. The stretching frequency of the adsorbate-substrate bond, ($\nu_{S-CO}$), is decreased by $\sim$ 10-20 cm$^{-1}$ compared to PW91. Hence, the DFT+U approach leads to comparable results as the calculations with RPBE, from 20 to 35 cm$^{-1}$ below the experimental values.

A stronger influence of the corrected HOMO--LUMO gap is found for the workfunction.
The shorter C-O bond resulting from the DFT+U calculations decreases the workfunction compared to a PW91 calculation due to the reduction of the dipole moment by $\sim$ 0.2-0.3 eV. The workfunction change of the system where the CO molecule sits on top of Cu becomes negative ($\Delta \Phi = $--0.22 eV) on the Cu(111) surface and slightly positive ($\Delta \Phi \sim$ 30 meV) for on-top CO on Cu(001).

We can conclude that the DFT+U method slightly modificates the structural and vibrational properties which are in better agreement with experiment than the values obtained by PW91 and RPBE exchange-correlation functional.

\section{Conclusion}

A detailed analysis of the DFT predictions for CO adsorption on the Cu(111) and Cu(001) surface was presented. Calculations with two different GGA functional produce rather accurate adsorption geometries and vibrational eigenfrequencies for CO adsorbed in the on-top position. However, using the PW91 functional too large adsorption energies and a wrong site preference are predicted on both surfaces. Replacing the PW91 by the RPBE functionals hardly changes the vibrational frequencies and the geometry (except the relaxation of the top layer) and results in adsorption energies in much better agreement with experiment. However, the site-preference is corrected only for the Cu(001), but not for the Cu(111) surface.

It was demonstrated that the site preference may be corrected using a molecular DFT+U approach which increases HOMO-LUMO separation in the CO molecule. This leads to a reduced electron donation into the antibonding $2\pi^\star$ orbital and reduces the strength of the chemisorption, the reduction being more pronounced for a higher coordination. The weaker adsorption strength results also in a higher stretching frequency of the adsorbed molecule, also improving agreement with experiment. Altogether we find that the molecular DFT+U method is a very convenient method to correct for the underestimation of the HOMO-LUMO gap characteristic for the DFT and for achieving a better description of the adsorption of difficult molecules such as CO.

\section*{Acknowledgements}
We thank Georg Kresse for communicating unpublished material and for enlightening discussions.


\end{document}